\documentclass{article}
\usepackage{srcltx}
\usepackage{amsmath}
\usepackage{amssymb}
\usepackage{amsfonts}
\usepackage{latexsym}
\usepackage{textcomp}
\usepackage{appendix}
\usepackage{multirow}
\usepackage{booktabs}
\usepackage{subfigure}
\usepackage{epsfig}
\usepackage{url}
\usepackage{array}
\usepackage{marvosym}
\usepackage{graphics}
\usepackage{graphicx}
\title{A permutation Information Theory tour through different interest rate maturities: \\
the Libor case}

\author{Aurelio Fern\'andez Bariviera\\ \scriptsize{Department of Business, Universitat Rovira i Virgili, Av. Universitat 1, 43204 Reus, Spain} \\ \scriptsize{\ttfamily aurelio.fernandez@urv.net}   \and M. Bel\'en Guercio \\  \scriptsize{Instituto de Investigaciones Econ\'omicas y Sociales del Sur, UNS-CONICET.} \\  \scriptsize{12 de Octubre y San Juan, B8000CTX Bah\'{\i}a Blanca, Argentina.} \\ \scriptsize{Universidad Provincial del  Sudoeste (UPSO).} \\ \scriptsize{ Alvarado 328, B8000CJH Bah\'ia Blanca, Argentina} \and Lisana B. Martinez \\   \scriptsize{Instituto de Investigaciones Econ\'omicas y Sociales del Sur, UNS-CONICET.} \\  \scriptsize{12 de Octubre y San Juan, B8000CTX Bah\'{\i}a Blanca, Argentina.} \\ \scriptsize{Universidad Provincial del  Sudoeste (UPSO).} \\ \scriptsize{ Alvarado 328, B8000CJH Bah\'ia Blanca, Argentina} \and Osvaldo A. Rosso \\ \scriptsize{Instituto de F\'{\i}sica, Universidade Federal de Alagoas (UFAL). } \\ 
\scriptsize{BR 104 Norte km 97, 57072-970 Macei\'o, Alagoas, Brazil. }\\ \scriptsize{Instituto Tecnol\'ogico de Buenos Aires (ITBA),} \\ \scriptsize{Av. Eduardo Madero 399, C1106ACD Ciudad Aut\'onoma de Buenos Aires, Argentina.}}

\begin{document}
\maketitle

\begin{abstract}

This paper analyzes Libor interest rates for seven different maturities and referred to operations 
in British Pounds, Euro, Swiss Francs and Japanese Yen, during the period  years 2001 to 2015. 
The analysis is performed by means of two quantifiers derived from Information Theory: the permutation 
Shannon entropy and the permutation Fisher information measure. 
An anomalous behavior in the Libor is detected in all currencies except Euro during the years 2006--2012. 
The stochastic switch is more severe in 1, 2 and 3 months maturities. Given the special mechanism of Libor 
setting, we conjecture that the behavior could have been produced by the manipulation that was uncovered by financial authorities.
We argue that our methodology is pertinent as a market overseeing instrument.


\end{abstract}%
%%%% I N T R O D U C T I O N %%%%%%%%%
\section{Introduction}
\label{sec:Intro}
Since the seminal work of Bachelier \cite{Bachelier1900}, prices in a competitive market have been modeled 
as a memoryless stochastic process. 
In fact, according to the Efficient Market Hypothesis (EMH), prices fully reflect all available information 
\cite{Fama76}. 
This property was duly proved by Samuelson \cite{Samuelson65}. 
It is known that informational efficiency can vary over time. 
This may be due to several reasons. 
For example, Cajueiro and Tabak \cite{Cajueiro2006} studied the Chinese stock market and found that liquidity  
plays a role in explaining the evolution of the long term memory; Bariviera \cite{Bariviera2011} argued that 
informational efficiency in the Thai stock market is influenced by liquidity constraints; while Bariviera 
{\it et al.\/} \cite{Bariviera2012} detailed the influence of the 2008 financial crisis on the memory endowment 
of the European fixed income market. 
However, changes in informational efficiency are essentially unpredictable, under the EMH framework. 

The problem addressed in this paper stems from a newspaper. 
Mollenkamp and Whitehouse \cite{Mollenkamp2008} published a disruptive article in the Wall Street 
Journal: they suggested that the Libor rate did not reflect what it was expected, \textit{i.e.}, the cost of funding of prime banks. 
The British Bankers Association (BBA) defines Libor as ``...the rate at which an individual Contributor 
Panel bank could borrow funds, were it to do so by asking for and then accepting inter-bank offers in 
reasonable market size, just prior to 11:00 [a.m.] London time''. 
Every London business day, each bank in the Contributor Panel (selected banks from BBA) makes a blind 
submission such that each banker does not know the quotes of the other bankers. 
A compiler, Thomson Reuters, then averages the second and third quartiles. 
This average is published and represents the Libor rate on a given day. 
In other words, \emph{Libor is a trimmed average of the expected borrowing rates of leading banks.\/}  
Libor rates has been published for ten currencies and fifteen maturities. 
As it is defined, Libor is expected to be the best self estimate of leading banks borrowing cost at 
different maturities.
Several publications in newspapers \cite{Saigol2013,Reuters2012} casting doubts on Libor integrity triggered 
investigations by several surveillance authorities such as the US Department of Justice, the UK Financial 
Services Authority or the European Commission. 
All these offices found several traces of misconduct, which resulted in severe fines to leading banks. 

There are a few papers dealing with this topic in academic journals. 
Most of them are focused, once the manipulation is known, on verifying or discarding its existence, by means 
of statistical tests. 
Taylor and Williams \cite{Taylor2009} documented the detachment of the Libor rate from other market 
rates such as Overnight Interest Swap (OIS), Effective Federal Fund (EFF),  Certificate of Deposits (CDs), 
Credit Default Swaps (CDS), and Repo rates. 
Snider and Youle \cite{Snider2010} studied individual quotes in the Libor bank panel and  found that Libor 
quotes in the US were not strongly related to other bank borrowing cost proxies. 
Abrantes-Metz {\it et al.\/} \cite{Abrantes2011} analyzed the distribution of the Second Digits (SDs) of 
daily Libor rates between 1987 and 2008 and, compared it with uniform and Benford's distributions. 
If we take into account the whole period, the null hypothesis that the empirical distribution follows either 
the uniform or Benford's distribution cannot be rejected. 
However, if we take into account only the period after the subprime crisis, the null hypothesis is rejected. 
This result calls into question the ``aseptic" setting of Libor.
Monticini and Thornton \cite{Monticini2013} found evidence of Libor under-reporting after analyzing the spread
between 1-month and 3-month Libor and the rate of Certificate of Deposits using the Bai and Perron 
\cite{Bai1998} test for multiple structural breaks. 
For a historical overview of the Libor case from a regulator' point of view, see the Federal Reserve Staff 
Report \cite{Hou2014}.

Recently, Bariviera {\it et al.\/} \cite{Bariviera2015A,Bariviera2015B} present preliminary results about 3-month UK Libor manipulation. They performed a symbolic time series analysis using the Complexity Entropy Causality Plane (CECP). Our approach goes deeper in two aspects. First, we study the behavior of the Libor for several maturities and currencies. Second, we introduce a local information theory quantifier, which is able to detect tiny perturbations in the probability density function. Consequently, instead of analyzing our results with global vs. global quantifiers, as in the CECP, we study the time series by means of global vs. local quantifiers, defined in the Shannon-Fisher plane. 

Interest rates have broad and wide economic impact on our daily life. 
The lack of integrity of Libor as an information signal gives market participants a wrong proxy of borrowing 
costs, thus providing a bad rate for pricing financial products. 
Many mortgages as well as sovereign bonds have Libor-linked interest rates. 
Therefore, the wrong setting of the interest rates produces a cascade effect throughout the whole economy. 
Consequently, Libor reliability is crucial to private and public borrowers  around the world.

Our approach is somewhat different to the previous literature. 
The aim of this paper is to propose a quantitative technique based on Information Theory quantifiers, in order 
to detect changes in the stochastic/chaotic underlying dynamics of the Libor time series. 
We claim that our methodology is able to capture changes in the process dynamics on an ongoing basis. 
In other words, our method allows periodic monitoring of interest time series. 
This market audit, done at regular intervals, can detect unusual mutation in the statistical properties of a 
time series. 
It is clear that not only manipulation can shift signal features. 
Other external forces, such as increasing noise or special economic situations influence on signal observation 
and measurement. 
Therefore, results should be read with care. 
A change in the stochastic dynamics should not be read exclusively as produced by manipulation. 
It could be due to spurious contamination of the signal, or liquidity constraints, etc. 
However, our method acts as an early warning mechanism to detect some kind of ``market distress". 
This study is relevant no only for researchers, but also for surveillance authorities who need an efficient 
market watch device in order to detect strange movements in key economic variables such as the Libor. 

This paper is structured as follows. 
Section \ref{sec:methodology} describes the methodology. 
Section \ref{sec:data} details the data used in this paper. 
Section \ref{sec:results} discuses our empirical findings. 
Finally, Section \ref{sec:conclusions} draws the main conclusions of our research.

%%%% INFORMATION THEORY QUANTIFIERS %%%%%%
\section{Methodology}
\label{sec:methodology}
Financial markets are complex, dynamic systems in which unobservable hidden structures govern their behavior. 
Usually, we can only observe an output, \textit{e.g.} an equilibrium price, an interest rate fixing, etc. 
As a consequence, the researcher should study the behavior of that output in order to infer the characteristics 
of the underlying dynamical phenomenon. 

Time series should be carefully analyzed in order to extract relevant information for simulation and forecasting 
purposes. 
Information Theory derived quantifiers can be considered good candidates for this task because they are able to 
characterize some properties of the probability distribution associated with the observable or measurable quantity.

If the Libor rates were somewhat manipulated (as suggested in Ref. \cite{Mollenkamp2008}), some change in 
the stochastic behavior should appear. 
To be more precise, according to the EMH, interest rate time series should behave approximately as an standard 
Brownian motion. 
Manipulation is, by definition, the introduction of an strange deterministic device into the time series. 
According to Wold \cite{Wold1938} a time series can be split into two components, one purely random and other 
deterministic. 
If manipulation is successful, the deterministic induced behavior should offset the random behavior. 
This process results in an reduction of the natural stochastic character of the interest rates. 
We argue that the selected Information Theory quantifiers are able to detect such reduction and its temporal 
duration.

We use two specific quantifiers: permutation Shannon entropy and permutation Fisher information measure. 
These quantifiers are evaluated in pairs, displaying them in $2D$-planar representation. 
This causal Shannon--Fisher Plane gives an insight on global versus local perturbations on the behavior of a time 
series of the underlying dynamics of a physical process under analysis.

\subsection{The Shannon entropy}
\label{sec:Shannon}
The Shannon entropy is usually regarded as a natural measure of the quantity of information in a physical process. 
Given a continuous probability distribution function (PDF) $f(x)$ with $x \in \Delta \subset {\mathbb R}$ and 
$\int_{\Delta} f(x)~dx = 1$, its associated {\it Shannon Entropy\/} $S$  \cite{Shannon1949} is
\begin{equation}
\label{shannon}
S[f]~=~-\int_{\Delta}~f~\ln(f)~dx \ ,
\end{equation}
a measure of ``global character"  that it is not too sensitive to strong changes in the distribution taking place 
on a small-sized region. 
Let now $P=\{p_i;~i=1,\cdots, N\}$  be a  discrete probability distribution, with $N$ the number of possible states 
of the system under study.
In the discrete case,  we define a ``normalized" Shannon entropy, $0 \leq {\mathcal H} \leq 1$, as
\begin{equation}
\label{shannon-disc}
{\mathcal H}[P]~=~ S[P]  / S_{max} ~=~\left\{-\sum_{i=1}^{N}~p_i~\ln( p_i) \right\} /  S_{max} \ ,
\end{equation}
where the denominator  $S_{max} = S[P_e] = \ln N$ is that attained by a uniform probability distribution 
$P_e = \{p_i =1/N,~ \forall i = 1, \cdots, N\}$.

\subsection{The Fisher information measure}
\label{sec:Fisher}
The {\it Fisher's Information Measure\/} (FIM) $\mathcal F$ \cite{Fisher1922,Frieden2004}, constitutes a 
measure of the gradient content of the distribution $f(x)$, thus being quite sensitive even to tiny localized 
perturbations. 
It reads
\begin{equation}
\label{fisher}
{\mathcal F}[f]~=~\int_{\Delta}~ { {1} \over {f(x)} } \left[ { {df(x)} \over {dx} }\right]^2 ~dx
    ~=~4 \int_{\Delta}~\left[ { {d \psi(x)} \over {dx} }\right]^2
\ .
\end{equation}
FIM can be variously interpreted as a measure of the ability to estimate a parameter, as the amount of information 
that can be extracted from a set of measurements, and also as a measure of the state of disorder of a system  or
phenomenon \cite{Frieden2004}.
In the previous definition of FIM (Eq. (\ref{fisher})) the division by $f(x)$ is not convenient if 
$f(x) \rightarrow 0$ at certain $x-$values. 
We avoid this if we work with a real probability amplitudes $f(x)= \psi^{2}(x)$ \cite{Fisher1922,Frieden2004},
which is a  simpler form (no divisors) and shows that $\mathcal F$ simply measures the gradient content in $\psi(x)$.
The gradient operator significantly influences the contribution of minute local $f-$variations to FIM's value.
Accordingly, this quantifier is called a ``local" one \cite{Frieden2004}.

Let $P=\{p_i;~i=1,\cdots, N\}$ be a  discrete probability distribution, with $N$ the number of possible 
states of the system under study. 
The concomitant  problem of information-loss due to discretization has been thoroughly studied and, in particular, 
it entails the loss of FIM's shift-invariance, which is of no importance for our present purposes 
\cite{Olivares2012A,Olivares2012B}.
For the FIM we take the expression in term of  real probability amplitudes as starting point, then a discrete 
normalized FIM,  $0 \leq {\mathcal F} \leq 1$, convenient for our present purposes, is given by
\begin{equation}
\label{Fisher-disc}
{\mathcal F}[P]~=~F_0~\sum_{i=1}^{N-1}~[(p_{i+1})^{1/2} - (p_{i})^{1/2}]^2 \ .
\end{equation}
It has been extensively discussed that this discretization is the best behaved in a  discrete environment 
\cite{Dehesa2009}. 
Here the normalization constant $F_0$ reads
\begin{equation}
\label{F0}
F_0~=~\left\{
       \begin{array}{cl}
                    1       &\qquad \mbox{if $p_{i^*} = 1$ for
                            $i^* = 1$ or $i^* = N$ and $p_{i}  = 0 ~\forall  i \neq i^*$} \\
                    1/2     &\qquad \mbox{otherwise}
       \end{array}
\right.
\ .
\end{equation}

If our system lies in a very ordered state, which occurs when almost all the $p_{i}$ -- values are zeros 
except for a particular state $k \neq i$ with $p_{k} \cong 1$, we have a normalized Shannon entropy 
${\mathcal H} \sim 0$ and a normalized Fisher's information measure ${\mathcal F} \sim 1$.
On the other hand, when the system under study is represented by  a very disordered state, that is when all 
the $p_{i}$ -- values oscillate around the same value we obtain ${\mathcal H} \sim 1$ while ${\mathcal F} \sim 0$.
One can state that the general FIM--behavior of the present discrete version (Eq. (\ref{Fisher-disc})),  
is opposite to that of the Shannon entropy, except for periodic motions \cite{Olivares2012A,Olivares2012B}.
The local sensitivity of FIM for discrete--PDFs is reflected in the fact that the specific ``$i-$ordering" 
of the discrete values $p_{i}$ must be seriously taken into account in evaluating the sum in Eq.~(\ref{Fisher-disc}).
This point was extensively discussed by Rosso {\it et al.\/} in previous works \cite{Olivares2012A,Olivares2012B}.
The summands can be regarded as a kind of ``distance" between  two contiguous probabilities.
Thus, a different ordering of the pertinent summands would lead to a different FIM-value, hereby its local nature.
In the present work, we follow the lexicographic order described by Lehmer \cite{Lehmer} in the generation 
of Bandt-Pompe PDF (see next section).
Given the local character of FIM, when combined with a global quantifier as the Shannon entropy, conforms the 
Shannon--Fisher plane, ${\mathcal H} \times {\mathcal F}$, introduced by Vignat and Bercher \cite{Vignat2003}. 
These authors showed that this plane is able to characterize the non-stationary behavior of a complex signal.

\subsection{The Bandt Pompe method for PDF evaluation}
\label{sec:BP}
Time series (TS) analysis, \textit{i.e.} temporal measurements of variables, is a very important area of economic science. 
In particular, it is very important to extract information in order to unveil the underlying dynamics of irregular 
and apparently unpredictable behavior in a given market. 
The starting point of TS analysis is to determine the most appropriate probability density function associated 
with the TS. 
Several methods compete for its proper estimation. 
For example Rosso {\it et al.\/} \cite{Rosso2009c} propose the frequency of occurrence; 
De Micco {\it et al.\/} \cite{Demicco2008} propose amplitude-based procedures; 
Mischaikow {\it et al.\/} \cite{Mischaikow1999} prefer binary symbolic dynamics; 
Powell {\it et al.\/} \cite{Powell1979} recommend Fourier analysis and;
Rosso {\it et al.\/} \cite{Rosso2001} introduce wavelets transformation, among others. 
The suitability of each of the proposed methodologies depends on the peculiarity of data, such as stationarity, 
length of the series, the variation of the parameters, the level of noise contamination, etc. 
In all these cases, global aspects of the dynamics can be somehow captured, but the different approaches are 
not equivalent in their ability to discern all relevant physical details.
Bandt and Pompe \cite{Bandt2002} introduced a simple and robust symbolic method that takes into account time 
order  of the TS. 

The pertinent symbolic data are:
{\it (i)\/} created by ranking the values of the series; and
{\it (ii)\/} defined by reordering the embedded data in ascending order, which is tantamount to a phase space 
reconstruction with embedding dimension (pattern length) $D$ and time lag $\tau$.
In this way, it is possible to quantify the diversity of the ordering symbols (patterns) derived from a scalar 
time series.
Note that the appropriate symbol sequence arises naturally from the time series, and no model-based assumptions 
are needed.
In fact, the necessary ``partitions'' are devised by comparing the order of neighboring relative values rather 
than by apportioning amplitudes according to different levels.
This technique, as opposed to most of those in current practice, takes into account the temporal structure of 
the time series generated by the physical process under study.
This feature allows us to uncover important details concerning the ordinal structure of the time series
\cite{Olivares2012B,Rosso2007,Rosso2012} and can also yield information about temporal correlation 
\cite{Rosso2009A,Rosso2009B}.

It is clear that this type of analysis of a time series entails losing some details of the original series' 
amplitude information.
Nevertheless, by just referring to the series' intrinsic structure, a meaningful difficulty reduction has indeed 
been achieved by Bandt and Pompe with regard to the description of complex systems.
The symbolic representation of time series by recourse to a comparison of consecutive ($\tau = 1$) or 
nonconsecutive ($\tau > 1$) values allows for an accurate empirical reconstruction of the underlying phase-space, 
even in the presence of weak (observational and dynamic) noise \cite{Bandt2002}.
Furthermore, the ordinal patterns associated with the PDF is invariant with respect to nonlinear monotonous 
transformations.
Accordingly, nonlinear drifts or scaling artificially introduced by a measurement device will not modify the 
estimation of quantifiers, a nice property if one deals with experimental data (see, \textit{e.g.}, \cite{Saco2010}).
These advantages make the Bandt and Pompe methodology more convenient than conventional methods based on range 
partitioning (\emph{i.e}., PDF based on histograms).

To use the Bandt and Pompe \cite{Bandt2002} methodology for evaluating the PDF, $P$, associated with the time 
series (dynamical system) under study, one starts by considering partitions of the pertinent $D$-dimensional 
space that will hopefully ``reveal'' relevant details of the ordinal structure of a given one-dimensional time 
series ${\mathcal X}(t) = \{ x_t; t= 1, \cdots, M\}$ with embedding dimension $D > 1$ ($D \in {\mathbb N}$) 
and embedding time delay $\tau$ ($\tau \in {\mathbb N}$).
We are interested in ``ordinal patterns'' of order (length) $D$ generated by
\begin{equation}
\label{asignation1}
(s)~\mapsto~ \left(~x_{s-(D-1)\tau},~x_{s-(D-2)\tau},~\cdots, \\ ~x_{s-\tau},~x_{s}~\right) \ ,
\end{equation}
which assigns to each time $s$ the $D$-dimensional vector of values at times $s, s-\tau,\cdots,s-(D-1)\tau$.
Clearly, the greater the $D-$value, the more information on the past is incorporated into our vectors.
By ``ordinal pattern'' related to the time $(s)$, we mean the permutation $\pi=(r_0,r_1, \cdots,r_{D-1})$ of
$[0,1,\cdots,D-1]$ defined by 
\begin{equation}
\label{asignation2}
x_{s-r_{D-1}\tau}~\le~x_{s-r_{D-2}\tau}~\le~\cdots~\le~x_{s-r_{1}\tau}~\le~x_{s-r_0\tau}  \ .
\end{equation}
In order to get a unique result, we set $r_i < r_{i-1}$ if $x_{s-r_{i}} = x_{s-r_{i-1}}$.
This is justified if the values of $x_t$ have a continuous distribution, so that equal values are very unusual.

For all the $D!$ possible orderings (permutations) $\pi_i$ when  embedding dimension is $D$, their associated 
relative frequencies can be naturally computed according to the number of times this particular order sequence 
is found in the time series,  divided by the total number of sequences,
\begin{equation}
\label{eq:frequ}
p(\pi_i)= \frac{\sharp \{s|s\leq N-(D-1)\tau ;~(s) \quad \texttt{has type}~\pi_i \}}{N-(D-1)\tau} \ .
\end{equation}
In the last expression the symbol $\sharp$ stands for ``number".
Thus, an ordinal pattern probability distribution $P = \{ p(\pi_i), i = 1, \cdots, D! \}$ is obtained from the 
time series.

Consequently, it is possible to quantify the diversity of the ordering symbols (patterns of length $D$) derived 
from a scalar time series, by evaluating the so-called permutation entropy (Shannon entropy) and  permutation 
Fisher information measure.
Of course, the embedding dimension $D$ plays an important role in the evaluation of the appropriate probability 
distribution, because $D$ determines the number of accessible states $D!$ and also conditions the minimum 
acceptable length $M \gg D!$ of the time series that one needs in order to work with reliable statistics 
\cite{Rosso2007}.

Regarding the selection of the parameters, Bandt and Pompe suggested working with $4 \leq D \leq 6$ and
specifically considered an embedding delay $\tau = 1$ in their cornerstone paper \cite{Bandt2002}.
Nevertheless, it is clear that other values of $\tau$ could provide additional information.
It has been recently shown that this parameter is strongly related, if it is relevant, to the intrinsic time 
scales of the system under analysis \cite{Zunino2010B,Soriano2011,Zunino2012}.

Additional advantages of the  method reside in
{\it i)\/} its simplicity (we need  few parameters: the pattern length/embedding dimension $D$ and the embedding 
delay $\tau$) and
{\it ii)\/} the extremely fast nature of the pertinent calculation-process \cite{Keller2005}.
The BP methodology can be applied not only  to time series representative of low dimensional dynamical systems 
but also to any type of time series (regular, chaotic, noisy, or reality based).
In fact, the existence of an attractor in the $D$-dimensional phase space in not assumed.
The only condition for the applicability of the BP method is  a very weak stationary assumption: for $k \leq D$, 
the probability for $x_t < x_{t+k}$ should not depend on $t$.
For a review of BP's methodology and its applications to physics, biomedical and econophysics signals see Zanin 
{\it et al.\/} \cite{Zanin2012}. 
Rosso \cite{Rosso2007} show that the above mentioned quantifiers produce better descriptions of the process 
associated dynamics when the PDF is computed using BP rather than using the usual histogram methodology.

The Bandt and Pompe proposal for associating probability distributions to time series (of an underlying symbolic 
nature) constitutes a significant advance in the study of nonlinear dynamical systems \cite{Bandt2002}.
The method provides univocal prescription for ordinary, global entropic quantifiers of the Shannon-kind.
However, as was shown by Rosso and coworkers \cite{Olivares2012A,Olivares2012B}, ambiguities arise in applying 
the Bandt and Pompe technique with reference to the permutation of ordinal patterns. 
This happens if one wishes to employ the BP-probability density to construct local entropic quantifiers, 
like the Fisher information measure, which would characterize time series generated by nonlinear dynamical systems.

The local sensitivity of the Fisher Information measure for discrete PDFs is reflected in the fact that the 
specific ``$i$-ordering'' of the discrete values $p_i$ must be seriously taken into account in evaluating the 
sum in Eq.~(\ref{Fisher-disc}).
The pertinent numerator can be regarded as a kind of ``distance'' between two contiguous probabilities.
Thus, a different ordering of the pertinent summands would lead to a different Fisher information value.
In fact, if we have a discrete PDF given by $P = \{ p_i, i = 1, \cdots , N\}$, we will have $N!$ possibilities
{for the $i$-ordering.}

The question is, which is the arrangement that one could regard as the ``proper'' ordering?
The answer is straightforward in some cases, the histogram-based PDF constituting a conspicuous example.
For such a procedure, one first divides the interval $[a, b]$ (with $a$ and $b$ the minimum and maximum
amplitude values in the time series) into a finite number on non-overlapping sub-intervals (bins).
Thus, the division procedure of the interval $[a, b]$ provides the natural order sequence for the evaluation
of the PDF gradient involved in the Fisher information measure.
In our current paper, we chosen for the Bandt--Pompe PDF the lexicographic ordering given by the algorithm
of Lehmer \cite{Lehmer}, amongst other possibilities, due to it provide a better distinction of different 
dynamics in the Shannon--Fisher plane, ${\mathcal H} \times {\mathcal F}$ (see \cite{Olivares2012A,Olivares2012B}).

%%%%%%%% D A T A %%%%%%%%%%%%
\section{Data}
\label{sec:data}
We analize the Libor rates in British Pounds (GBP), Euro (EUR), Swiss Franc (CHF) and Japanese Yen (JPY), 
for the following seven maturities: overnight (O/N), one week (1W), one month (1M), two months (2M), three 
months (3M), six months (6M) and twelve months (12M). 
The data coverage is from 02/01/2001 until 24/03/2015, for a total of 3711 data points. 
All data were retrieved from Datastream.  

%%%%%%% R E S U L T S %%%%%%%%%%
\section{Results}
\label{sec:results}
We analyze each time series using the methodology described in Section \ref{sec:methodology}. 
The PDF was computed following the BP recipe, because it is the single method to introduce time-causality 
in PDF building. 
Although we have performed our analysis using embedding dimensions $D = 3,4$ and 5 with time lag $\tau =1$, 
we present the results only for $D = 4$, because it exhibits the best clarity for our explanations. 
The other embedding dimensions results are similar. 

%We produce two types of graphs. 
%One is the Complexity Entropy Causality Plane CECP), that shows the global behavior of the time series. 
%The other is the Shannon--Fisher Plane (SFP), which combines the global and local dimensions in the scrutiny 
%of the time series. 
%Taking into account that the discrete version of FIM (displayed in Equation \ref{eq:discreteFIM}) can be 
%regarded as a kind of ``distance'' between two contiguous probabilities, different ordering of the pertinent 
%summands would lead to a different FIM-value. 
%Although, this is the essence of its local nature, it requires the appropriate selection of a ``natural'' 
%order. 
%There are several order candidates. 
%Our experience suggest that a lexicographic order known as Lehmer \cite{Lehmer} results in a good choice 
%to discriminate dynamics. 

We would like to know if the underlying generating process changes during the observation period. 
In order to evaluate such change, we construct sliding windows. 
The sliding windows work as follows: 
We consider the first $N = 300$ data points and evaluate the two quantifiers,  ${\mathcal H}$ and ${\mathcal F}$. 
Then we move $\delta = 20$ data points forward and compute the quantifiers of the new window with the following 
$N = 300$ data points. 
We continue this process until the end of the time series. 
After doing so, we obtain 170 estimation windows. 
Each window spans approximately one year, and the window step is approximately one month. 

The initial and final dates of each estimation window is displayed in %Table 1 \ref{tab:period1} and Table \ref{tab:period2}
the electronic supplementary material ESM-Table 1.

We  also present in the electronic supplementary material the graphs corresponding to  
${\mathcal H} \times {\mathcal F}$-plane for the Libor rates in GBP (ESM-Fig.~1), Euro (ESM-Fig.~2),
CHF (ESM-Fig.~3), JPY (ESM-Fig.~4) for the seven considered maturities: O/N, 1W, 2M, 3M, 6M and 12M, when 
they are evaluated on the 170 estimation windows respectively, given in this way a representation of 
the time evolution of them.
From these figures, it can be observed that 1M, 2M and 3M maturities have similar shape.
It can also noted that 6M and 12M are very similar to each other and their localization points are more
concentrated.
We can summarize these results as: two distinct dynamics are detected. 
On one side, 1M, 2M and 3M maturities scatters throughout the plane. 
On the other side, O/N, 1W, 6M and 12M graphs are less disperse and their points are more concentrated on 
the low right corner. 
In the financial economics vocabulary, we can say that the latter maturities are more informationally 
efficient: their time series are closer to a random walk. 
The middle maturities (1M, 2M, 3M) occupy places in the ${\mathcal H} \times {\mathcal F}$--plane that are 
consistent with fractional Brownian motion or with chaotic dynamics. 

This is the first evidence that we are dealing with a single interest rate but with distinct generating 
processes. 
In order to highlight the unequal behavior between the two groups of maturities, we count the number of 
points of the ${\mathcal H} \times {\mathcal F}$--plane that are closer to the low right corner and 
the number of points that are away from that corner. 
We divide the points into two regions. 
One, the most informational efficient region comprises the points where ${\mathcal H} > 0.75$ and
${\mathcal F} < 0.3$. 
The other is the complement of the ${\mathcal H} \times {\mathcal F}$--plane. 
We display the results of the point counting in Tables \ref{tab:pointsGBP} to \ref{tab:pointsJPY}.
These tables confirm the ocular inspection of the ${\mathcal H} \times {\mathcal F}$--plane
(see figures ESM-Fig.~1 to ESM-Fig.~4). 
Whereas for 1M, 2M and 3M maturities only between 33\% to 39\% of the points are in the efficient area, 
the percentages for the other maturities ranges 55\% to 63\%.  
We recall that each point reflects the estimation of the quantifiers at a given moving window. 

\begin{figure}[!h]
\centering
\includegraphics[scale=0.35]{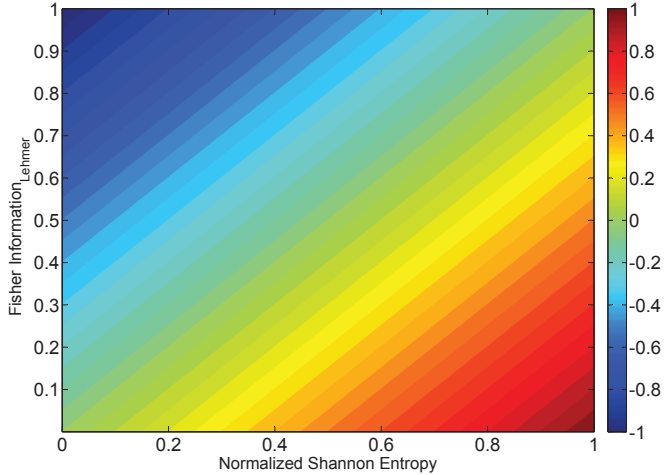}
\caption{
Color map that reflects the different degrees of efficiency in the Shannon--Fisher plane according to 
Eq.~(\ref{eq:efficiency}).
}
\label{fig:colormap}
\end{figure}

We would like to enquire if the aforementioned characteristics changed in an ordered way through time. 
Taking into account that we are studying the changes across time in the degree of informational efficiency 
of financial time series, we propose the following efficiency index, ${\mathcal E}$:
\begin{equation} 
\label{eq:efficiency}
{\mathcal E}[P] = {\mathcal H}[P] - {\mathcal F}[P] 
\end{equation}
Given that both quantifiers, ${\mathcal H}$ and ${\mathcal F}$, are bounded between 0 and 1, 
and that their behavior is expected opposite, the maximum value of our efficiency index is 
${\mathcal E}[P] = 1$, when ${\mathcal H}[P]=1 $ and 
${\mathcal F}[P]=0 $ and ${\mathcal E}[P] = -1$ when  ${\mathcal H} [P]=0$ and ${\mathcal F}[P]=1$ 
The most informational efficient behavior (\textit{i.e.} the most random behavior) is when Shannon entropy is maximized 
and Fisher information minimized. 
Consequently, our efficiency index ${\mathcal E}$ lies in the $[-1,1]$ interval. 
Accordingly, our efficiency index can be represented in a color map as in Fig.~\ref{fig:colormap}

%%%COLORMAP

We aim to assess the evolution of the efficiency ${\mathcal E}$ through time. 
In order to do so, we display the result for each currency in a color map where the $x$-axis is the temporal 
dimension (\textit{i.e.} the estimation windows) and the $y$-axis are the different currencies. 
The colors of the map reflects a degree of efficiency according to the color scale on the right of each graph 
and coincides with the color scale of Fig.~\ref{fig:colormap}.

\begin{figure}[!h]
    \centering
{
     \subfigure[GBP]{%
            \label{fig:colormapGBP}
            \includegraphics[width=0.8\textwidth]{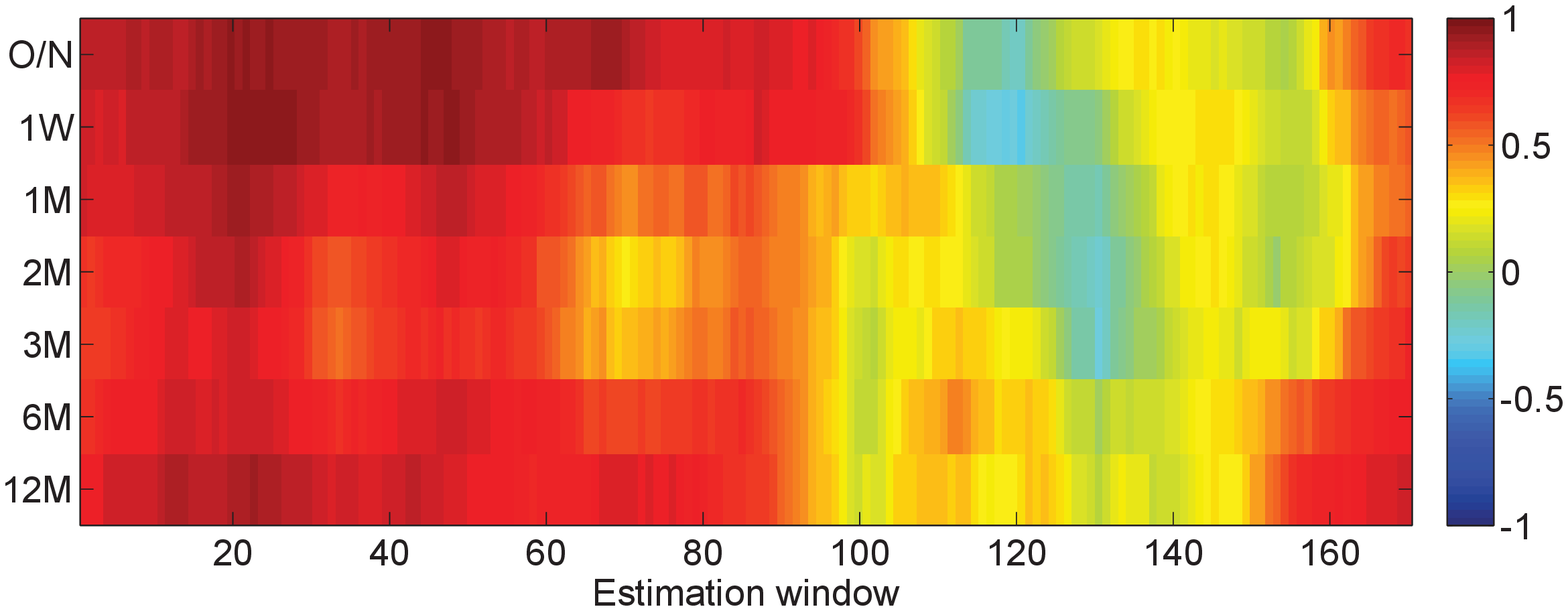}
        }%
        \\%  ------- End of the first row ----------------------%
     \subfigure[EUR]{%
          \label{fig:colormapEUR}
          \includegraphics[width=0.8\textwidth]{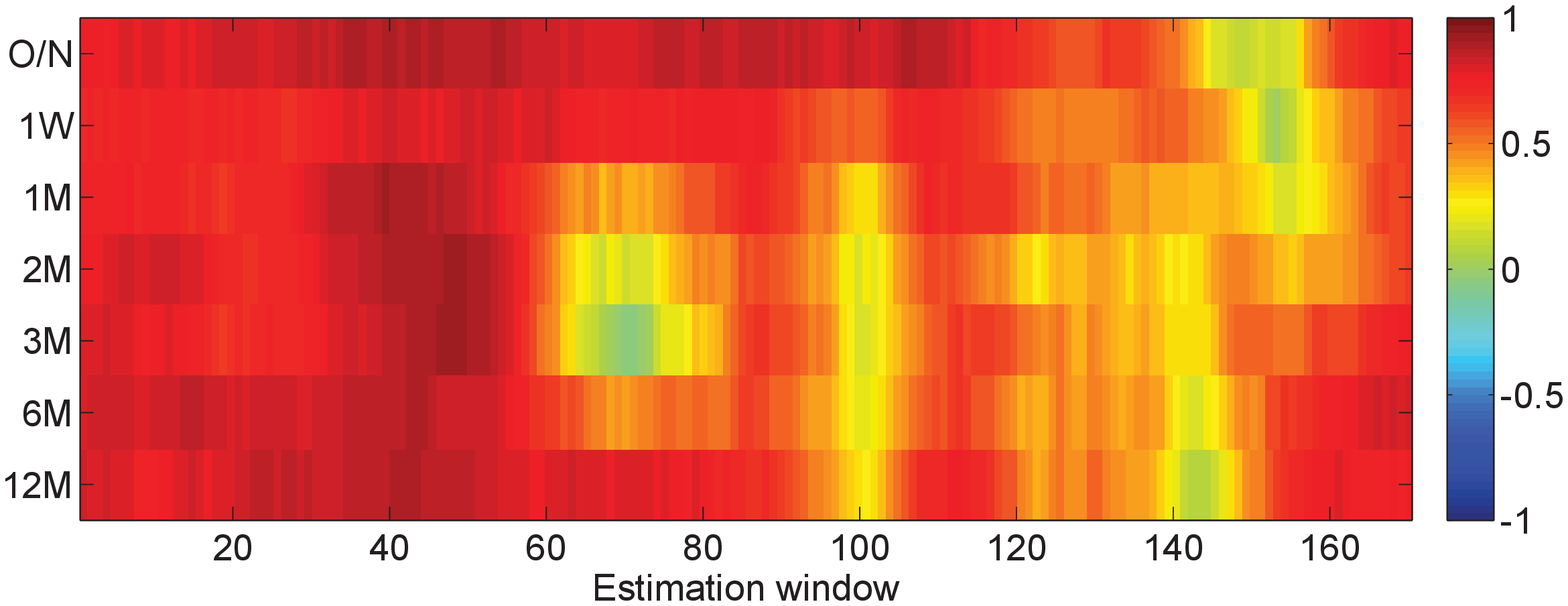}
        }
         \\%------- End of the second row ----------------------%  
     \subfigure[CHF]{%
          \label{fig:colormapCHF}
          \includegraphics[width=0.8\textwidth]{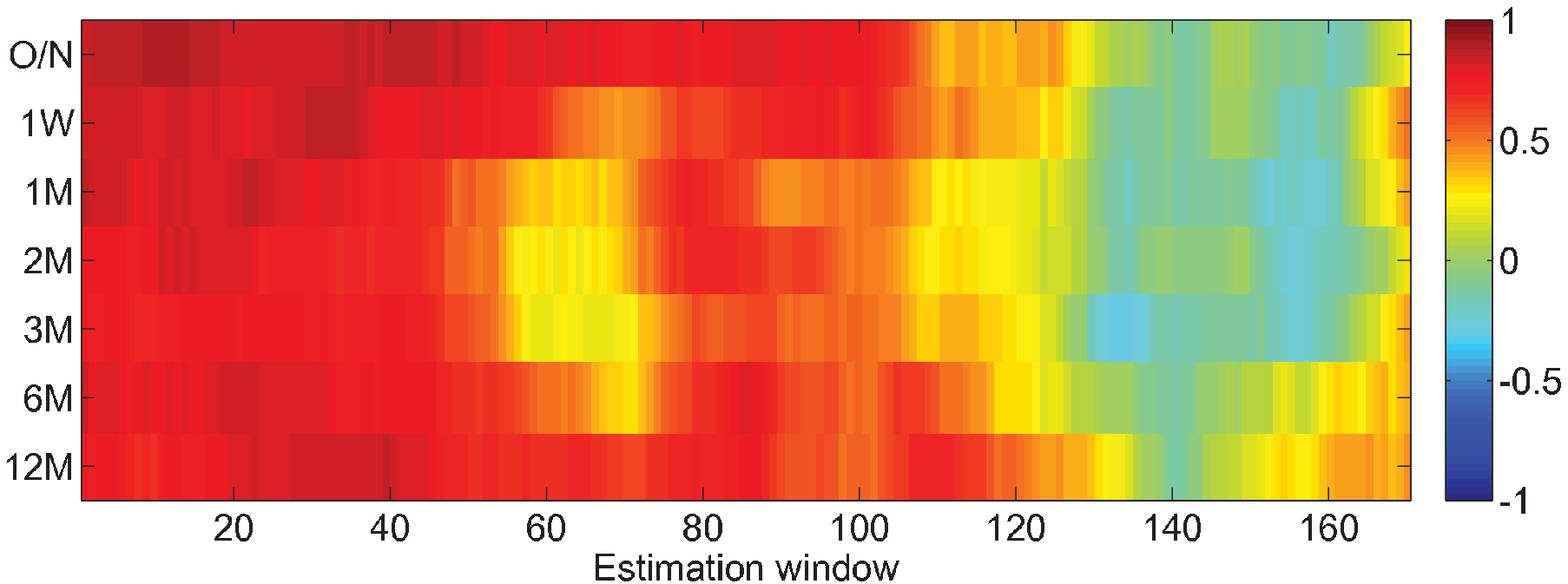}
       }
          \\%------- End of the third row ----------------------%   
    \subfigure[JPY]{%
          \label{fig:colormapJPY}
          \includegraphics[width=0.8\textwidth]{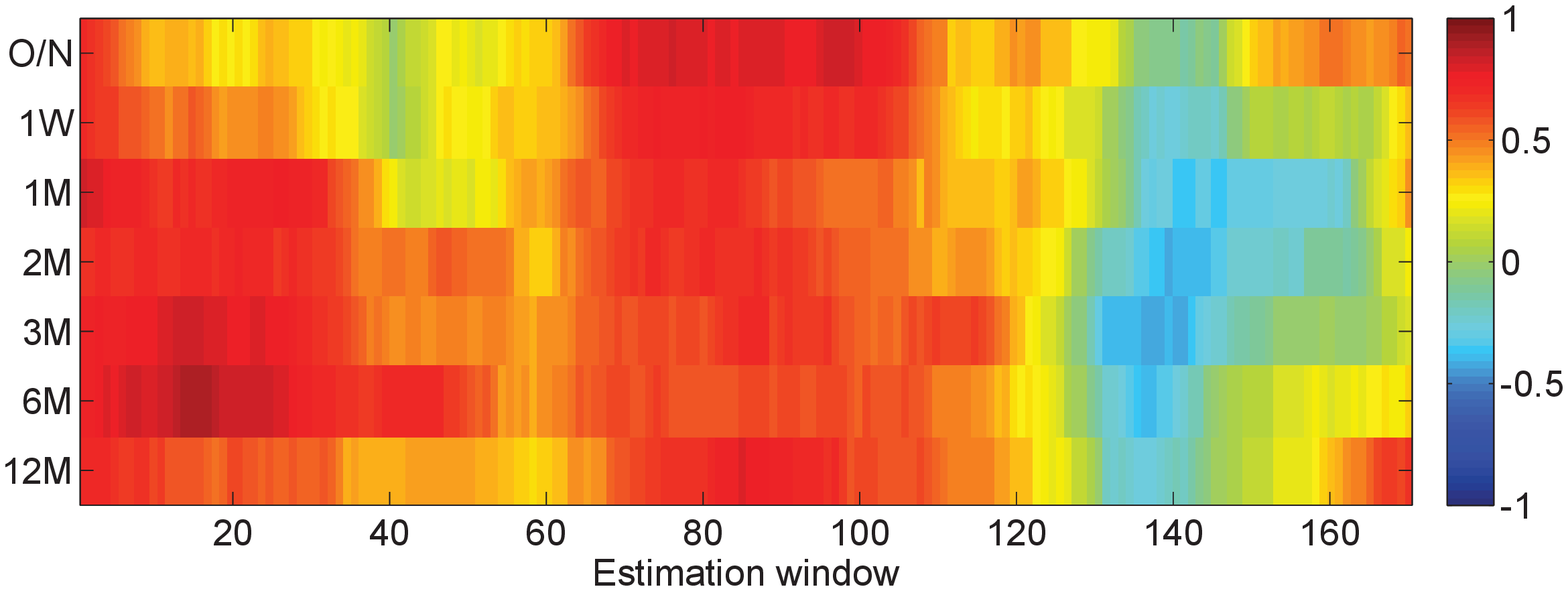}
      }
}
    \caption{Color map of the evolution of the degree of efficiency of different maturities and currencies of Libor. 
Efficiency is computed according to Eq. (\ref{eq:efficiency})
The information quantifiers were computing using Bandt-Pompe PDF with $D=4$ and $\tau = 1$.
    }%
   \label{fig:colormapsALL}
\end{figure}

The map corresponding GBP (see Fig.~\ref{fig:colormapGBP}) clearly reflects two features. 
First that O/N, 1W, 6M and 12M maturities has been behaving better that the middle maturities. 
Even more, 6M and 12M has been globally the most efficient during all the period.
On contrary, 1M, 2M, 3M were informationally efficient until window 100 (05/08/2008 to 31/08/2009) and suffered 
a sudden shift in its stochastic behavior, that returns to its previous path around window 160 
(12/03/2013 to 05/05/2014). 
Additionally we can observe a yellow area around windows 65--80, that roughly correspond to years 2006 and 2007. 
The behavior detected by this color map is consistent with the alleged manipulation of Libor rates reported in 
the press \cite{Mollenkamp2008}. 
According to our results, Libor rates suffered from some kind of inefficient behavior (coherent with manipulation) 
during the period 2006--2012. 
In year 2013 the levels of informational efficiency recovered to levels similar to periods previous to the financial 
crisis.

\begin{table}[htbp]
  \centering
  \caption{
Number of points in the inefficient (${\mathcal H} < 0.75,~{\mathcal F} > 0.3$) and 
efficient (${\mathcal H} > 0.75,~{\mathcal F} < 0.3$) regions for different maturities of Libor GBP.
The information quantifiers were computing using Bandt-Pompe PDF with $D=4$ and $\tau = 1$.}
    \begin{tabular}{rrrr}
    \toprule
		Points  within  &  Points outside    &      Percentage of & Series \\
              efficiency bounds &  efficiency bounds &  efficient windows &  \\
    \midrule   
    107   & 63    & \textbf{63\%} &  GBP  Libor O/N \\
    103   & 67    & \textbf{61\%} &  GBP  Libor 1W \\
    67    & 103   & 39\%          &  GBP  Libor 1M \\ 
    56    & 114   & 33\%          &  GBP  Libor 2M \\
    58    & 112   & 34\%          &  GBP  Libor 3M \\
    93    & 77    & \textbf{55\%} &  GBP  Libor 6M \\
   106    & 64    & \textbf{62\%} &  GBP  Libor 12M \\
    \bottomrule
    \end{tabular}%
  \label{tab:pointsGBP}%
\end{table}%

The Euro market exhibits a similar pattern with respect to the two groups of maturities. 
However, the loss of efficiency is was less severe. 
This can be observed in the Fig.~\ref{fig:colormapEUR}, where there is a prevalence of orange and red colors.

%%%EURO 
\begin{table}[htbp]
  \centering
  \caption{
Number of points in the inefficient (${\mathcal H} < 0.75,~{\mathcal F} > 0.3$) and 
efficient (${\mathcal H} > 0.75,~{\mathcal F} < 0.3$) regions for different maturities of Libor EUR.
The information quantifiers were computing using Bandt-Pompe PDF with $D=4$ and $\tau = 1$.}
    \begin{tabular}{rrrr}
    \toprule
        Points within  &  Points outside    & Percentage of     & Series \\
     efficiency bounds &  efficiency bounds & efficient windows &  \\
         \midrule
    145   & 25    & \textbf{85\%} &  EUR  Libor O/N \\
    113   & 57    & \textbf{66\%} &  EUR  Libor 1W \\
    87    & 83    & 51\%          &  EUR  Libor 1M \\
    68    & 102   & 40\%          &  EUR  Libor 2M \\
    84    & 86    & 49\%          &  EUR  Libor 3M \\
    90    & 80    & \textbf{53\%} &  EUR  Libor 6M \\
    121   & 49    & \textbf{71\%} &  EUR  Libor 12M \\
    \bottomrule
    \end{tabular}%
  \label{tab:pointsEUR}%
\end{table}%

The Swiss Franc market behavior (see Fig.~\ref{fig:colormapCHF}) is more similar to the British Pound market
(see Fig.~\ref{fig:colormapGBP}) 
The region that comprises windows 120 until 160 reflects an eroded informational efficiency, that was slowly 
recovered at the final of the observation period. 
Additionally there is a yellow spot around windows 60--70 exclusively in 1M, 2M and 3M maturities. 

%%% CHF
\begin{table}[htbp]
  \centering
  \caption{
Number of points in the inefficient (${\mathcal H} < 0.75,~{\mathcal F} > 0.3$) and 
efficient (${\mathcal H} > 0.75,~{\mathcal F} < 0.3$) regions for different maturities of Libor CHF.
The information quantifiers were computing using Bandt-Pompe PDF with $D=4$ and $\tau = 1$.}
    \begin{tabular}{rrrr}
    \toprule
    Points within      & Points outside     & Percentage of     & Series \\ 
     efficiency bounds &  efficiency bounds & efficient windows &  \\
       \midrule
    107   & 63    & \textbf{63\%} & CHF  Libor O/N \\
    95    & 75    & \textbf{56\%} & CHF  Libor 1W \\
    59    & 111   & 35\%          & CHF  Libor 1M \\
    66    & 104   & 39\%          & CHF  Libor 2M \\
    58    & 112   & 34\%          & CHF  Libor 3M \\
    78    & 92    & \textbf{46\%} & CHF  Libor 6M \\
    102   & 68    & \textbf{60\%} & CHF  Libor 12M \\
    \bottomrule
    \end{tabular}%
  \label{tab:pointsCHF}%
\end{table}%

The Japanese Yen market (see Fig.~\ref{fig:colormapJPY}) was also affected in the degree of efficiency. 
In general, we can observe in the color map that this is a less informational efficient market: light orange, 
yellow and light blue are dominant. 
There is also a severe erosion of the informational efficiency between windows 120 to 160, recovered at the 
end of the observation period mainly in O/N and 12M maturities. 

%%% JPY
\begin{table}[htbp]
  \centering
  \caption{
Number of points in the inefficient (${\mathcal H} < 0.75,~{\mathcal F} > 0.3$) and 
efficient (${\mathcal H} > 0.75,~{\mathcal F} < 0.3$) regions for different maturities of Libor JPY.
The information quantifiers were computing using Bandt-Pompe PDF with $D=4$ and $\tau = 1$.}
    \begin{tabular}{rrrr}
    \toprule
    Points within  & Points outside & Percentage of  & Series \\
     efficiency bounds &  efficiency bounds & efficient windows &  \\
         \midrule
    49    & 121   & 29\%          & JPY  Libor S/N \\
    50    & 120   & 29\%          & JPY  Libor 1W \\
    59    & 111   & \textbf{35\%} & JPY  Libor 1M \\
    64    & 106   & \textbf{38\%} & JPY  Libor 2M \\
    64    & 106   & \textbf{38\%} & JPY  Libor 3M \\
    82    & 88    & \textbf{48\%} & JPY  Libor 6M \\
    68    & 102   & \textbf{40\%} & JPY  Libor 12M \\
    \bottomrule
    \end{tabular}%
  \label{tab:pointsJPY}%
\end{table}%

%%%%%%%%%%C O N C L U S I O N S %%%%%%%%%%%
\section{Conclusion}
\label{sec:conclusions}
The first finding that we can extract from our data analysis is that the Libor for all currencies and for maturities 1M, 2M and 3M is governed by a different process than the other maturities. The mentioned maturities usually reflect a lower level of informational efficiency than the other maturities. The second important finding is that during the years 2007-2012 (windows 120-160) there was a significant reduction in the informational efficiency of all markets, except the Euro market. This behavior seems to be contemporary to the uncovered manipulation announced in the newspapers \cite{Mollenkamp2008}.
We would like to highlight that our method is not intended to find manipulation, but rather to uncover changes in the hidden stochastic structure of data. However, it is able to clearly discriminate areas of more deterministic behavior in the time series, which could be consistent with the alleged interest rate rigging. 

\vskip6pt

%\disclaimer{Insert disclaimer text here.}

%\bibliographystyle{vancouver} %rspublicnat
%\bibliography{liborbib}       % name your BibTeX data base

%%%%%%%%%% Insert bibliography here %%%%%%%%%%%%%%

\end{document}